# Formalising Surveillance and Identity

**Victoria Wang & John V. Tucker**

**Abstract**

Surveillance is a social phenomenon that is general and commonplace, employed by governments, companies and communities. Its ubiquity is due to technologies for gathering and processing data; its strong and obvious effects raise difficult social questions. We give a general definition of surveillance that captures the notion in diverse situations and we illustrate it with some disparate examples. A most important, if neglected, component idea is that of the identity of the people or objects observed. We propose a general definition of identifiers as data designed to specify the identity of an entity in some context or for some purpose. We examine the ways identifiers depend upon other identifiers and show the provenance of identifiers requires reductions between identifiers and a special idea of personal identifier. The theory is formalised mathematically. Finally, we reflect on the role of formal methods to give insights in sociological contexts.

**Keywords:** surveillance, identity, identifier, identity tree, reductions, streams, abstract data type, surveillance systems, identity management systems, formal methods.

## 1. Introduction

Surveillance is an integral part of everyday life as many technologies employed in our physical and virtual environments are capable of monitoring and recording. Ubiquitous cameras that monitor our physical environment to improve the safety and security of people and property are but the most visible tip of the surveillance iceberg. The invisible bulk is made of data and software. Our professional lives have long been conducted through software systems that record data about actions and events. Recently, our personal lives have become dependent on software systems too. There is the sociological phenomenon of the representation of life in data. Technologically, there is the translation and unification of all kinds of data into digital forms,which is combined with the transfer and unification of all kinds of sources of data brought about by the connectivity of computer networks.

Surveillance is enormously controversial as it impacts on the multitude of notions that make up privacy for individuals; the conduct ofsocial and economic life of societies; and on the legal, political, and military foundations of the state. With this broad view, David Lyon has given a general description of surveillance as "the focused, systematic and routine attention to personal details for purposes of influence, management, protection or detection" (2007: 14). As a general social issue, Lyon has proposed that surveillance has three main purposes: (i) *keeping control*, which is the historic purpose pursued by employers, police, government; (ii) *social sorting*, pursued by companies in marketing and managing customers; and (iii) *mutual monitoring*, pursued peer to peer in social networks, real and virtual (Lyon, 2007; our own italics). Of these, the methods and aims of social sorting seem to becoming universal.

In this paper we examine theoretically the general idea of surveillance and one of its component concepts, that of identity. We seek completely abstract models that can be formalised and analysed mathematically. First, we develop a general definition of surveillance that captures the notion in diverse situations and illustrate it with some disparate examples. The definition shows that the three main purposes of surveillance have very similar structures and suggest new examples of surveillance. Our analysis applies to objects, real or virtual, as well as people.

A most important component idea of surveillance is that of the identity of the people or objects observed. We introduce the general concept of *identifiers,* which are defined to be data designed to specify the identity of an entity in some context or for some purpose. Identifiers are the focus of our analysis and as a starting point for our theory we propose that:

**Principle.** *Entities are "known" only through the data that act as their identifiers.*

This idea seems to be valid in general, even historically; the identity of a person is reduced to forms of evidence including personal testimony and biometrics. In our time, the idea is simpler and more palatable when one considers the virtual world, since it is created and constantly changing. Users have many identities, which they create in a state of anonymity. Technically, the operations and tests on identifiers combine to make systems that are specific to some context or task, though with unforeseen applications. Since identifiers are data, clearly the systems of identifiers are what computer scientists call abstract data types.

Foremost among identifiers are those that are supposed to identify people. The notion of a personal identifier proves to be as informative as it is subtle. To understand identity we need to examine the ways identifiers are issued and how they depend upon other identifiers. We show that the provenance of identifiers is an essential idea. We consider principles of how identifiers are to be compared and when they might be deemed equivalent; this requires notions of reductions between identifiers.

All of these ideas are motivated by informal described examples and then formalised mathematically using elementary algebra and logic. In their mathematical form, the theoretical notions are precise and reveal most clearly the possible structure of ideas.

Our aim is to clarify concepts and their interdependencies from which structure emerges. Our formalisation can be likened to the way formal logic has long been used in philosophy to clarify the nature of arguments and reasoning. Formal logic has proved to a fundamental *practical* tool in the development of computer programming and software engineering. Whether the formal models here are useful to technologists designing surveillance systems – or their safeguards – remains to be seen. Finally, we reflect on the role of abstract concepts and formal methods to give insights in sociological contexts.

## 2. What is Surveillance?

Let us begin with an abstract description of a large class of surveillance systems.

***Informal Definition.*** A *surveillance system* observes the behaviour of people and objects in space and time; it classifies behaviours into attributes; and it identifies people and objects with some of those attributes. A surveillance system consists of the following components and methods:

1. ***Entity.*** Entities that are people or objects that possess behaviour in space and time;
2. ***Observable behaviour.*** Methods for observing and recording behaviours;
3. ***Attribute.*** Methods for defining and recognising attributes of behaviours, based on rules, norms, practices, expectations, and other observable properties; and
4. ***Identity.*** Methods for generating data that identifies entities that exhibit the attributes and locate them in space and time.

Although we may expect the attributes to be deviations from sets of rules, norms, practices, etc., the definition does *not* require or imply deviance. The definition does require precise formulations of attributes. The data that is used to identify entities are invariably numbers and texts, but could be sounds and images. Here are two simple examples to prepare for our formalisations.

***Example 1: Control – Motor Vehicles.*** Automatic Number Plate Recognition (ANPR) is a technology that observes vehicles and recognises number plates or registration marks, possibly using infrared so as to function day and night. Common applications are checking on vehicle speed, managing car parking and collecting tolls. The technology was functioning in the late 1970s; today, ANPR can be found in thousands of fixed surveillance systems owned by both public and private organisations. We describe some ANPR applications in terms of our abstract definition.

The entities in such surveillance systems are cars at a particular location and time; they may be in transit (speed check), or entering or leaving a location (car park, congestion charge). The method the system uses for observing the cars is a camera that creates an image that may be communicated and stored. This image is processed by software that will recognise a behavioural attribute (e.g., breaking a speed limit) and, in particular, performs optical character recognition to establish the registration mark of a car. The registration mark is an alpha-numeric name that identifies a vehicle uniquely. On communicating this registration mark, the identity of the entity is established. For example, the output of such surveillance systems is the identity of a vehicle travelling too fast, or arriving or leaving a particular location. Alternately, and in summary, a surveillance system for car parking based on an ANPR is:

*Entity*: Cars

*Observable Behaviour*: Time of arrival and departure at location

*Attributes*: Duration of stay above a particular limit

*Identity*: Registration marks

Actually, such surveillance systems are normally thought to be observing drivers. In this case, we take the entities to be people. Following the ANPR stages described above, the registration mark is communicated to a database relevant to the application. For example, the database may be used to check an attribute, such as a payment (tax, charge or toll) having been made for that registration mark. The surveillance system knows the identity of the car, but not necessarily the driver.

In order to find the driver an independent process involving only identity begins. The keeper of the vehicle must be located and contacted. Suppose the operator of the surveillance system communicates the registration mark with the Driver and Vehicle Licensing Agency (DVLA) to determine the name and address of the keeper. The output of these actions is the identity of the keeper – finding the actual driver may require further independent action (that is not a part of the surveillance system). Note that in this stage there is a transformation of identity data from the registration mark to the name and address of the keeper.

***Example 2: Social Sorting – Customer Accounts.*** Consider a client's account with some service provider, such as a bank, insurance company or shop. Commonly, such an account has the following basic structure (See: Figure 2). There is a user name and password that act as a key simply to gain access to the account. The account details establish basic information such as: name, address, services provided, etc. The behaviour is the account history that not only records the past transactions but allows all sorts of new transactions, queries, preferences, etc. to be performed. It is the account history that is clearly subjected to tests that seek, for example, that terms and conditions are met by the client or that no unusual pattern of transactions has been carried out. In summary:

*Entities:* Credit card accounts

*Observable Behaviour*: Transactions: date, payee, location, sum, etc.

*Attributes:* Credit limit, minimum payments, unusual transactions

*Identity:* Credit card number

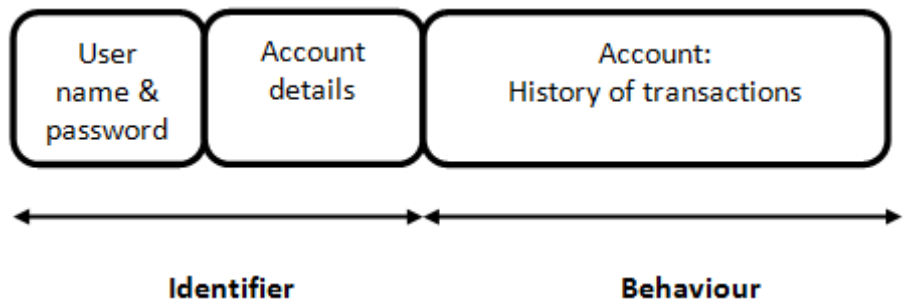

Figure 1: A Typical Customer Account

***Example 3: Mutual Monitoring – Social Media Accounts***

Social media connect people who have personal or professional interests in common. Systems such as Facebook, Twitter, LinkedIn and Academia.edu attract large numbers of users: individuals register with a system and create an account and a network of other users to suit their needs. Abstractly, the account has a structure similar to that of a customer account for a bank, insurance company or shop (see: Figure 2). The behaviour of the account is a history of status updates, linkages and interactions. Such social networking is firmly based on the fact that individuals voluntarily reveal very detailed information about themselves, their tastes and opinions, and their activities to their networks. From the point of view of surveillance, two phenomena are of interest: (i) individuals can and do 'watch over' the people in their networks, and (ii) all the data on all of the account holders belong to companies that can collect and use the information for commercial purposes. In summary:

*Entities*: Accounts

*Observable Behaviour*: Tweet

*Attributes*: Keyword

*Identity*: Usernames

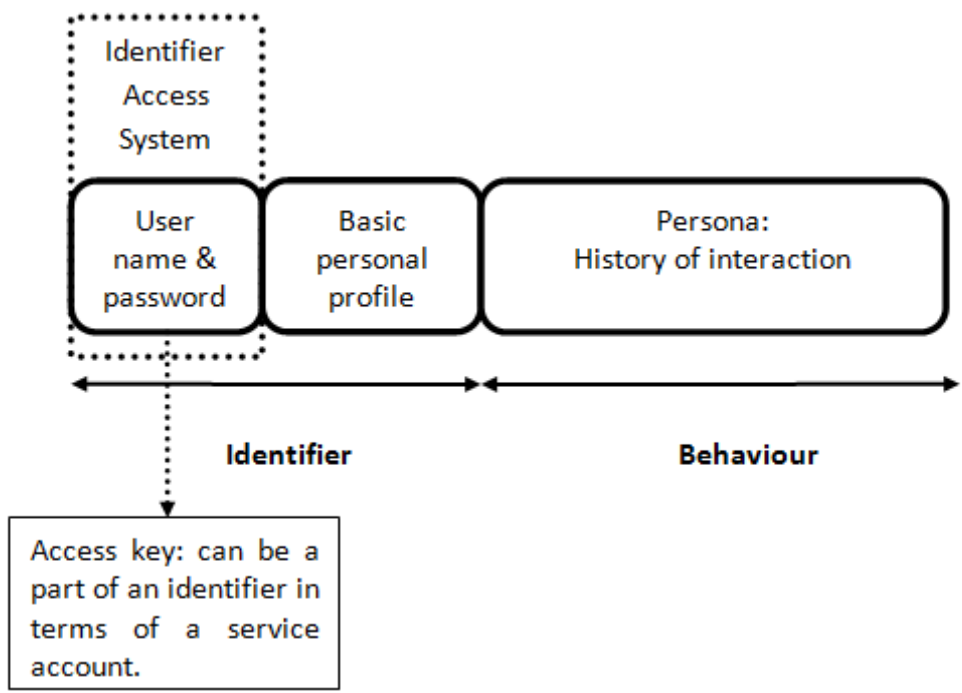

Figure 2: A Typical Media Account

## 3. A Formal Model of Surveillance

We have defined surveillance as a process that observes entities seeking to detect some property of their behaviour and that identifies entities with the property. In this section we will redefine this notion of surveillance formally. We will use elementary set theory[1] to create a precise and general definition that covers a great range of examples.

### 3.1 Entities and their behaviour

Entities may be people or objects, real or virtual. Let *Entity* be a set of entities whose behaviour is to be observed. Let *Behaviour* be the set of all *possible* behaviours in space and timeof the all the entities of *Entity*. The nature of behaviour and its models we will consider later.

---

[1]The mathematical ideas we use are sets, functions and relations, which are described in many textbooks on discrete mathematics, e.g., Lipschutz and Lipson (2009), Makinson (2012).

For simplicity, we suppose that each entity $e \in Entity$ has *one and only one* behaviour in space and time, i.e., its behaviour is deterministic. In this case, there is a single-valued mapping

$$[[\_]]: Entity \rightarrow Behaviour$$

such that

$$[[e]] = \text{the behaviour of the entity } e \in Entity.$$

The mapping provides a formal meaning or semantics for the behaviour of the entity.

To define surveillance formally, next we have to establish what it is we are to detect. The purpose of surveillance is to detect some property of the behaviour of an entity – e.g., some action or event. Commonly, what is of interest is a deviation from the some standards of behaviour of the entity. To formalise the target property we simply suppose that *Prop* is some property of behaviours, i.e.,

$$Prop \subseteq Behaviour.$$

The elements of *Behaviour* formalise the activity of the entities. The entities of interest are those whose behaviour lie in *Prop*; in symbols:

$$Prop\text{-}Entity = \{e \in Entity : [[e]] \in Prop\}$$

The behaviours need to be modelled formally. How is this done?

## 3.2. Behaviour as streams of data

Typically, entities behave in time and their activities involve actions or events of some kind. A way to formalise behaviours is to think of a sequence of actions or events taking place in time. Let us show how these can be defined formally as streams, which are sequences of data indexed by time.

Let *Time* be a set of time points; for example, say $Time = \{0, 1, 2, \ldots, t, \ldots\}$. The time points are data generated by a clock of some kind.

Let *Action* be a set of actions or events characteristic of the entities. The behaviour of an entity is conceived of as a sequence of actions or events in time:

$$a(0), a(1), a(2), \ldots, a(t), \ldots$$

where these a($t$)∈*Action* for all $t$∈*Time*. Such sequences can be usefully termed *traces* or *histories.*

**Definition.** A *trace* is an association of actions or events to time points and is formalised by a mapping

$$a: Time \rightarrow Action$$

such that for all $t$∈*Time*

$a(t)$ = the action or event taking place at time $t$∈*Time*.

Let *Trace* be the set of all possible traces.

Now in many cases, the space *Behaviour* of all possible behaviours of the entities can be taken to be a subset of the set *Trace* of all possible traces; thus,

$$Prop \subseteq Behaviour \subseteq Trace$$

When applying the behaviour mapping [[_]] to an entity $e$∈*Entity* we get a trace, which is a map

$$[[e]]: Time \rightarrow Action.$$

Therefore, for $e$∈*Entity* and $t$∈*Time*, we have

[[$e$]]($t$) = the action or event of entity $e$ taking place at time $t$∈*Time*.

### *Example. Twitter*

Twitter processes data called *tweets*. At the heart of a tweet is a simple message made from at most 140 characters, but a tweet is composed of much more data. For simplicity, a tweet can be thought of as a vector of data drawn from sets of the following kind:

*Text*          The text that is the status update (using UTF-8 representation for Unicode standard)

*Identity*       A string that uniquely labels the tweet

*Contributor*   The author(s) of the tweet.

*Time*         The time when this Tweet was created (measured by Coordinated Universal Time (UTC).

| *Location* | The geographic location (longitude, latitude) of this Tweet as reported by the user or application (using geoJSON standard). |
| --- | --- |
| *Retweet* | Status and number of retweets |
| *Favourite* | Number of favourites |

We let the set of all possible tweets be

$$Tweet = Text \times Identity \times Contributor \times Time \times Location \times Retweet \times Favourite$$

Now Twitter generates and processes streams of tweets, i.e., sequences of tweets indexed by time. Thus, the behaviour can be modelled by traces that are streams of tweets of the form

$$a(0), a(1), a(2), \dots, a(t), \dots \in Tweet,$$

which is represented by a map

$$a: Time \rightarrow Tweet.$$

Let *Behaviour* be the set of all possible traces of these kinds. Typical user operations on tweets, once created, can be *embedding tweets*, *responding to tweets*, and *favouring, unfavouring*, and *deleting tweets*. These operations are operations on streams.

Depending upon the circumstances, monitoring tweet feeds is called *curation, filtering,* or *surveillance.* Monitoring Twitter can be done in a number of ways via Application Programming Interfaces (APIs), which define instructions for developers to build new systems. Twitter's *Search API* allows users to define criteria (keywords, usernames, locations, named places, etc) to search among existing tweets. Twitter's *Streaming API* redirects a *sample* of tweets, based upon a user's criteria, as these tweets appear. The sample is less than 1% (Morstatter et al., 2013). Twitter's *Firehose API* delivers 100% of all publicly available tweets that match users' criteria as they are made. The Twitter Firehose is complex and requires a subscritption. It is handled by two data providers, GNIP and DataSift. Twitter's monitoring services have police tools built upon them.

---

## 3.3 Identifying entities

The purpose of surveillance is to detect and identify entities whose behaviours lie in the subset *Prop*. Each entity $e \in Entity$ has some datum that is used to attempt to identify the entity. We will call this datum an *identifier*.[2] The association of identifiers with entities can become complicated as we will see in the following sections. At this stage, it is sufficient to suppose that identifiers have been assigned to entities

**Definition.** Let *Identifier* be a set of possible identifiers for the entities of *Entity*. There is a relation

$$id \subseteq Identifier \times Entity$$

such that

$id(i, e) \Leftrightarrow$ the data $i \in Identifier$, called an identifier, is assigned to entity $e \in E$.

Let *anon* be a datum that is *not* in the set *Identifier* of identifiers for the entities.

In our general setting, the association of identifiers to entities is a relation $id \subseteq Identifier \times Entity$ and so many identifiers can be allocated to many entities. A simple formulation of surveillance is to find *at least one* identifier for any entity whose behaviour satisfies *Prop*. This view of surveillance is embodied in the following mapping: define

$$Surv(Prop): Entity \rightarrow Identifier \cup \{anon\}$$

For $e \in Entity$ by

$$Surv(Prop)(e) = (some\ i)\ id(i, e) \qquad \text{if } [[e]] \in Prop$$
$$= anon \qquad\qquad\qquad \text{if } [[e]] \notin Prop.$$

Note that entities whose behaviours do not lie in *Prop* are ignored and will not be identified, i.e., they will remain anonymous. Combining these ideas we can define formally a very general notion of a surveillance system.

---

[2] In computing, the term *identifier* is well established and is data that names a computational component; invariably, it is a string that defines a component in a programming language such as locations, procedures, program etc. Our use is a generalisation and is close to the idea of a pure name in Needham (1989). The term is in use in some discussions of identity.

**Definition.** A *surveillance system* for a property of entities in a context has the form

$$SurvSys = (Entity, Identifier, Behaviour, Prop \mid id, [[\_]], Sur(Prop)),$$

consisting of the four non-empty sets

$$Entity, Identifier, Behaviour, Prop$$

and the two mappings

$$[[\_]]: Entity \rightarrow Behaviour$$

$$Surv(Prop): Entity \rightarrow Identifier \cup \{anon\},$$

and the relation

$$Id \subseteq Identifier \times Entity.$$

The definition proposes a *general logical form* of a surveillance system. It is a *formal specification* that establishes precisely the essential components of the system. Users and designers should have these in mind *before* thinking of any specific technological implementation. Any actual surveillance system will contain many technologies. These technologies will suggest some new logical components that can be explored in theorising and formalising implementations.

**Design Problem.** For a user and designer the essence of their cooperation is:

1.  *Specification*. To define the desired surveillance system:

    $$(Entity, Identifier, Behaviour, Prop \mid id, [[\_]], Surv(Prop)).$$

2.  *Implementation*. To choose technologies to

    a.  represent the behaviours of the entities;

    b.  observe behaviours and recognise those behaviours having the property;

    c.  represent the identities;

    d.  recognise the identity of entities

## 4. Surveillance and Social Sorting

The formal model of surveillance we have defined can be extended to include the phenomenon called *social sorting*. Social sorting is the categorization of populations and results in a classification used to treat people differently. The notion is central in surveillance. Although formulated to understand the social impact of surveillance on people, social sorting makes sense for entities of all kinds. We will formalise social sorting using more abstract notions of *categorization* and *partition*; we argue that it is rather problematic from a technical point of view because of the role of identity.

## 4.1. Sorting entities

At first sight, social sorting aims to produce a categorization of entities (especially people). What is a categorization?

**Definition.** Let *Entity* be a set of entities. A *categorization of entities* is a collection of subsets

$$S_1, S_2, \ldots, S_k \subseteq Entity$$

that include all the entities, i.e.,

$$S_1 \cup S_2 \cup \ldots \cup S_k = Entity$$

An entity *e* lies in at least one of the sets and possibly several. In this loose idea we may have categories overlapping and having interesting internal structure (e.g., they may form a hierarchy). Commonly, and most simply, we may want the sets not to overlap.

**Definition.** The classification is a *partition* if for $1 \leq n, m \leq k$, we have $S_n \cap S_m = \varnothing$.

## 4.2 Sorting identifiers

However, our understanding of surveillance observes behaviours of entities – not entities – and recognises only identifiers for entities – not the entities themselves. Thus, surveillance for sorting delivers a categorization of identifiers not entities. This makes the notion quite subtle.

**Definition.** Let *Identifier* be a set of identifiers. A *categorization of identifiers* is a collection of subsets

$$S_1, S_2, \ldots, S_k \subseteq Identifier$$

that includes all the identifiers, i.e.,

$$S_1 \cup S_2 \cup \ldots \cup S_k = Identifier$$

Again, an identifier $i$ lies in at least one of the sets and possibly several. However, the structure must be measured against the entities that the identifiers name. Given an entity $e$ there can be identifiers $i$ and $j$ for $e$ that lie in *different* classes. This means that the categorization of identifiers does *not* lead to a categorization of entities. There are ambiguities that need to be studied. Ideally, our categorization of identifiers can be transformed into one that corresponds with the entities:

**Definition.** The categorization *respects* the entities if for all $i, j \in Identifier,$

If $i \in S_n$ and $i$ and $j$ name the same entity then $j \in S_n$

Questions of ambiguity arise that can be resolved the equality of identifiers.

## 5. What is Identity?

Essential to establishing identity is the collection, storage and processing of data. Indeed, identity is almost purely a matter of data. People and objects are named, labelled or otherwise denoted by data relevant to contexts. The data in question captures some relevant aspects of a person or an object. Different identities are managed by different kinds of identity management systems. Be it physical or virtual, an identity is presented by a data type

We will look at some examples of *identity management systems* prior to providing a working definition, covering birth certificate, health records, driving license and National Insurance (NI) number, to demonstrate that identifiers are composite objects, in the sense that they are built from other identifiers.

*Basic personal identifiers* are those upon which we rely upon to distinguish a unique human being that is their guarantee of identity or be it to some contexts with their own level of rigor.

Thus, the purpose of an identifier is to establish when entities are the same or not in the surveillance context. Identifiers need not reflect any aspect of the entity or have any meaning at all. In our conception of surveillance, entities are observed and identified. This means that necessarily, surveillance systems must have methods to define the identity of entities. An identifier for an entity is a name that is associated to the entity and no other identity.

**Working Definition.** An *identifier* for an entity is data that is associated with the entity for the purposes of identifying it among similar entities.

For example, a name for an entity is an identifier. By a name for an entity we commonly mean data made from symbols. In terms of symbols, usually, numbers are added to identifiers in order to make an identifier unique to the entity.

The relationship between entities and identifiers can be complicated. Consider these four situations:

1. ***Many – One Associations***. Different identifiers can be assigned to the same entity.
2. ***One – One Associations***. Different identifiers are assigned to different entities.
3. ***One – Many Associations***. An identifier can be assigned to more than one entity but an entity has only one identifier.
4. ***Many – Many Associations***. An identifier is assigned to more than one entity and, vice versa, an entity can be assigned more than one identifier.

Surveillance reports identifiers that can narrow the search for entities but need not pin down the particular entity of interest. Thus, one-to-one associations are important because:

**Search Principle:** *If an association is many-one then given an identifier, we can search for a set of entities with that identifier.*

**Uniqueness Principle:** *If an association is one-one then given an identifier, we can search for the unique entity with that identifier.*

The following point is obvious but certainly profoundly important practically:

**Enumeration Principle:** *The addition of a number to an identifier of an entity can turn any many-one association into a one-one association.*

We will illustrate some of these with example, to which we will return in the formal theory.

***Example 1: Cars.*** This example illustrates one-one and many-one associations. Each car is assigned a registration mark, commonly known as registration number. The current system of UK was introduced on 1[st] September 2001. In general, each registration mark consists of seven characters with a defined format. From left to right, the characters consist of: (i) a local memory tag or area code, consisting of two letters that indicates the local registration office; (ii) a two-digit age identifier, which changes twice a year, in March and September; and (iii) a three-letter sequence which uniquely distinguishes each of the cars displaying the same initial four-character area and age sequence. The association of registration marks to cars is one-one. A car has one and only one registered keeper. Thus, the association of a registration marks to a keeper is unique. However, a person can be a registered keeper of as many cars as he/she wants. Thus, the association of registration marks to keepers is many-one.

The registration document (V5) for a car identifies the car and its keeper. However, it is not proof of ownership. The registered keeper is the person who is legally responsible for the car and need not to be the owner of the vehicle. Many people have insurance policies that enable them to drive any car with the owner's permission. Thus, the driver of a car on a particular occasion may be only loosely connected to the keeper. The association of cars to drivers is one-many. In terms of formal documents (containing several identifiers), the association between registration marks and drivers is complicated and probably incomplete.

***Example 2: Communications.*** This example demonstrates both many-one and many-many associations. When connecting a computer to the Internet, a number is needed called an IP address (32 bits under IP Protocol 4) that uniquely identifies the machine in the network. In some computer networks, such as networks local to an organisation or company, there is an IP address for the machine that does not change; these are called static IP addresses. The association of computers to IP addresses is one-one. More commonly, at home IP addresses are generated by the Internet Service Provider in response to a customer's need for Internet access. Thus, overtime IP addresses can change and the association of IP addresses to a particular computer is many-one. Developing this example, if more than one computer is accessing the Internet at the same time in a period, from the same service then the association between IP addresses and computers is many-many. The changing status seems to be natural in time-dependent associations of identifiers.

*Example 3: Addresses.* This example demonstrates a one-many association. In the UK a system of postal codes was introduced, between 1959-1974, to enable the automation of postal services. Typically, each address or location is assigned at most one postcode but a postcode can be assigned to more than one unit or building. The association between postcodes and buildings/addresses is one-many.

Thus, postcodes are a system of identifiers that do not uniquely determine addresses. Postcodes have found many uses and are used routinely in commercial transactions, navigation, and, more significantly, in calculating insurance, designing social policy and funding, and academic social studies – all of which are examples of social sorting.

*Working Definition.* An *identity management system* for a set of entities is a system with the following two properties:

    (i)    *Identifier Generation*: the system can create and delete identifiers for entities; and

    (ii)    *Identifier Authentication*: the system can, given two identifiers, decide whether or not they are associated with the same entity.

Another stronger formulation of authentication, which focuses on entities and the identifiers, is the following: *Entity Authentication*: the system can, given an entity and identifier, decide whether or not the identifier is associated with the entity.[3] The notion is attractive but problematic for what does it mean to be "given an entity"? In much theory and practice, the entity is actually given by means of another identifier!

## 6. A Formal Model of Identity

We now consider formally the idea of a system of identifiers for the entities under observation. Systems of identifiers can have many properties that require technical analysis, classification and application. There are three aspects arising from our discussion of examples: *assigning identifiers*, *comparing identifiers* and basic *personal identifiers*. We will continue to use the formal notations introduced earlier in our formal definition of surveillance in Section 3.

---

[3] Property (ii) is implied by this property.

**6.1 Assigning Identifiers**.

An identifier is some datum used to identify an entity in a context.

**Definition.** Suppose that identifiers have been assigned to entities and there is a relation

$$id \subseteq Identifier \times Entity$$

such that

$id(i, e)$ = the data $i \in I$, called an identifier, is assigned to entity $e \in E$.

We define the entities named by identifier $i$ by

$$ent(i) = \{ \, e \in Entity \mid id(i, e) \, \}$$

and all the identifiers naming entity $e$ by

$$id(e) = \{ \, i \in Identifier \mid id(i, e) \, \}.$$

These sets are projections of the relation *id*.

These are loose associations, wherein many identifiers can be associated with entities or many entities can have the same identifier. Let us make this idea our most general definition.

**Definition.** A *system of identifiers* is a structure consisting of two non-empty sets and a relation:

$$IdSys = (Identifier, Entity \mid id \subseteq Identifier \times Entity).$$

The examples suggest that the following special case is important.

***Example: Post Codes.*** Typically, each address/building is assigned at most one postcode but a postcode can be assigned to more than one unit or building. A one-many relations

$$code: Postcode \times Address.$$

**Definition.** A system of identifiers *IdSys* is said to satisfy the *many-one property* if each identifier is assigned to one entity but an entity may have many identifiers. In this case, the relation becomes a single-valued mapping

$$id: Identifier \rightarrow Entity$$

such that

$$id(i) = \text{the entity } e \in Entity \text{ named by the data } i \in Identifier.$$

Since the purpose of the identifiers is to recognise the entities that we are interested in, the following equivalence relation on *Identifier* is basic: for any $i_1$ and $i_2 \in Identifier$, we say they are *id-equivalent* if they are associated with the same entity: in symbols,

$$i_1 \approx_{en} i_2 \text{ if, and only if, } id(i_1) = id(i_2).$$

The identifier captures and narrows down detection of entities. Thus, we can strengthen the system of identifiers if we can satisfy this condition:

**Definition.** A system of identifiers *IdSys* is said to satisfy the one-to-one *uniqueness property* if entities have one and only one identifier and so the *identifier is unique to the entity*. In this case the map *id* satisfies this property: for any $i_1$ and $i_2 \in Identifier$,

$$\text{if } id(i_1) = id(i_2) \text{ then } i_1 = i_2.$$

The map *id* is *one-to-one* or *injective*.

***Example: Cars.*** Each car is assigned a registration mark, commonly known as registration number. The association of registration marks to cars is one-one.

## 6.2 Generating identifiers

How is an identifier generated for an entity in an identity management system? A few general ideas can be formulated. First, some input data is presented to the system that has to be examined and approved.

**Definition.** The data presented to a system to create an identifier is a specification of the entity. We will call this data a *form* and we let *Form* be the set of all possible forms for the system.

We represent the processing of the form by a function

$$Check: Forms \rightarrow \{0, 1\}$$

that tests the data in a form $f \in Form$ for consistency against the system's rules. We assume that *check(f) = 1* means the form is accepted and *check(f) = 0* means the form is rejected.

We represent the next stage – if and when an identifier is issued – by a function

$$issue: Forms \rightarrow Identifier$$

which uses some or all of the data in $f \in Form$ to make an identifier.

These stages are represented by composing the functions to make

$$generate: Forms \rightarrow Identifier$$

where

$$generate(f) = \textbf{if } check(f)=1 \textbf{ then } issue\ (f) \textbf{ else } reject$$

## 7. Personal Identity

Of greatest interest are surveillance systems in which the entities are people. A fundamental problem is how identifiers can actually identify a specific individual. We consider some examples of assigning data to individuals.

### 7.1 Examples

#### *Example 1: Biometrics*

Biometric identifiers are measurable qualities that can be used to describe and label the physical characteristics of individuals. Physiological characteristics are related to the body, and include fingerprints, photographs, palm prints, hand geometries, iris and retina images, odour/scent, and DNA. Behavioural characteristics are related to the behaviour of a person, including typing rhythm, gait and voice. The association of a biometric to people is expected to be one-one.

The operational tests used to measure biometrics, such as DNA, fingerprint and iris, are of course, approximate, due to technological constraints. Thus, that data presented manifests a one-one identity association is a matter of probability, especially high probability. Current studies suggest that increasingly accurate measurements can reveal differences in DNA between twins. Thus, although identical twins share very similar DNA, they are not identical (O'Connor, 2008). Identity is the subject of genetic research. Recently, public attention was drawn to this point when identical twins are identified by DNA evidence as suspects in a series of sexual assaults, in Marseille, France, and soon after in Reading, England. News reporting of the incidents is somewhat confused and incomplete. In the case of Marseille, it was reported that officials may have to pay about $1.3 million to compare billions pairs of nucleotides that make up DNA rather than compare 400 base pairs in a normal analysis (BBC, 2013). In the case of Reading, a commercial breakthrough in forensics was reported (Knapton, 2013).

***Example 2: Citizenship***

In the UK, for example, an individual can or must register with state organisations devoted to for health, employment, passport, and transport. Everyone registered with the National Health Service has his/her unique number, which is linked to his/her health record; each NHS number is made up of 10 alpha-numerics. Everyone gets a National Insurance (NI) number just before he/she turns 16. An individual's NI number makes sure his/her NI contributions and taxes are only recorded against her/her name. The format of the number is two prefix letters, six digits, and one suffix letter. In the new style red passport book, in addition to the biometrics, there is a passport number that must be nine characters and all characters must be numeric. Finally, each driving licence has a number made up of 18 alpha-numerics, which codes part or all of (i) the surname; (ii) the date of birth; (iii) the first names; (iv) sex; (v) licence issue; (vi) checks. In these cases of registration, numbers are added to identifiers in order to ensure each of these associations one-one.

## 7.2 Formal Person Identifiers.

A person identifier is data, but very special data. We wrote extensively about person identifiers earlier, citing the fundamental role of passports, National Insurance numbers, National Health Service numbers, and for some categories of people national identity cards.

We have emphasised how systems of identity are designed for certain purposes and that they are established with widely varying standards of rigour and are combined and compared in all sorts of ways. The fundamental person identifiers above carry weight: on the authority of the state identifying people and their basic situations for citizenship, employment, tax, and health.

**Definition.** A *personal identity system* has the form

$$PIdSys = (Identifier,\ Person\ |\ pid:\ Identifier \rightarrow Person).$$

and satisfies the uniqueness property, namely *two different people are assigned different data*.

In practice, the data assigned to a person invariably includes a number or alpha-numeric code precisely in order to *enforce* the uniqueness property.

The role of all systems of identity is understood by studying comparisons that involve reductions, but this is especially true of personal identity systems.

## 8. Provenance of Identifiers

### 8.1 Generating identifiers using other identifiers

Creating identifiers is an everyday occurrence; we open accounts, register for services, apply for permissions, buy products, etc. For many of these actions, we rely on a handful of pre-existing identifiers. To open a bank account, we give a proof of our identity and our current address, e.g., using a passport and a recent utility bill. To buy a product, an address and a credit card account number are usually sufficient for the vendor: notice the dependency on the bank identifier. At face value, the quality of a bank identifier is guaranteed by the databases of the state (passport, driver's licence) and, say, an energy provider (utility bill). The passport provides a high quality identifier based on a birth certificate, a photograph and possibly other biometric data. Example by example, illustrates the general point that:

**Principle.** *New identifiers are created from pre-existing identifiers*.

The dependability of one identifier upon another may be illustrated in an *identity tree* (see: Figure 3). The identifiers that appear in the nodes of the tree can create quite complicated dependency networks of identifiers.

The quality of an identifier is essentially a matter of its reliability, which in turn depends on its provenance, i.e., the process involved in establishing the identifier. In the case of people, a passport is a standard example of a high quality identifier with a rigorous provenance. In the case of a bank, the process is weaker as it is a now routine act of checking on identity.

Since identifiers are often built from other identifiers, of central importance is the process of comparing identifiers and reducing one type of identifier to another. Recognising a number plate of a car behaving badly can lead to a letter arriving at the address of the keeper and involves the transformation of a number of high quality identifiers (e.g., registration mark, keeper's name, address and driving history). All of these observations and ideas are formalised to make a precise and general mathematical framework for analysing identifiers.

In Figure 3 below, establishing the identifier ID1 involves providing evidence in the form of other identifiers: ID2-ID6. Thus, the validity of ID1 depends upon, or is reduced to, the validities of ID2-ID6. Some of these identifiers have a special status, in that they are designed to reliably denote an individual. In the example, these personal identifiers are guaranteed by the state (ID4) and biometric data (ID3); in the latter case, ID6 is used to allow a passport to be issued by post, without face-to-face interaction.

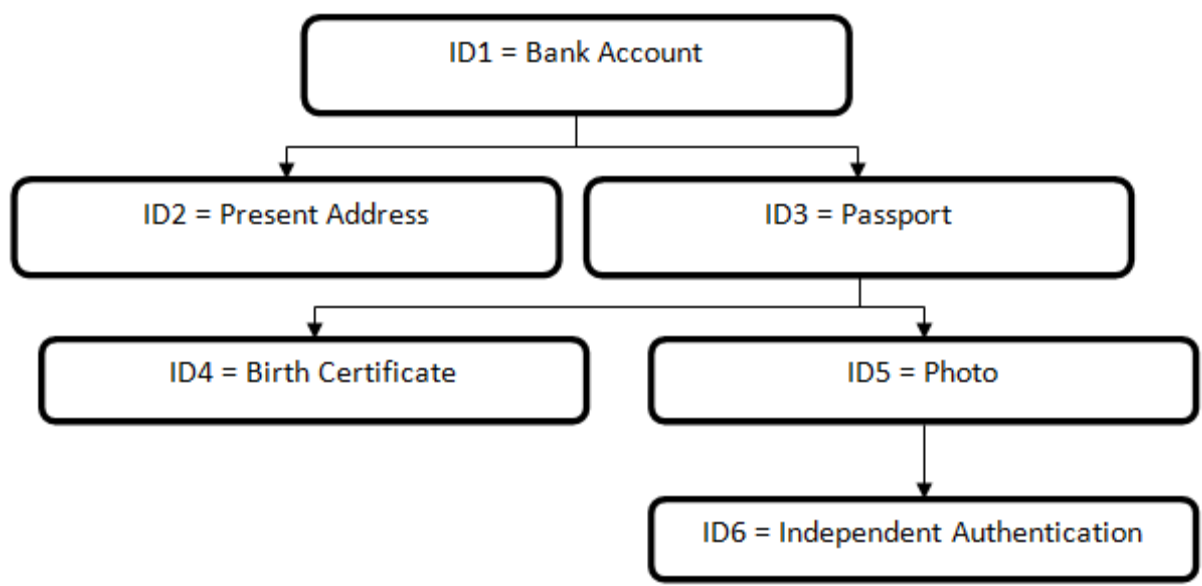

Figure 3: Identity Tree

## 8.2. Generating identifiers from identifiers

Now suppose that an identifier is generated for an entity in an identity management system. The few general ideas of Section 6.2 can be reformulated. Again, some input data is presented to the system that has to be examined and approved, but this time other identifiers must also be presented.

We represent the processing of the form by a function with new variables:

$$Check: Forms \times Identifier_1 \times ... \times Identifier_k \rightarrow \{0, 1\}$$

that tests the data in a form $f \in Form$ and the information available from identifiers $i_1, ... , i_k$ for consistency against the system's rules. As usual, we assume that $check(f, i_1, ... , i_k) = 0$ means the form is rejected and $check(f, i_1, ... , i_k) = 1$ means the form is accepted.

Let us assume the following:

**Provenance Principle.** *The data in f $\in$ Form is sufficient to create an identifier and that the data in the identifiers $i_1, ... , i_k$ are used only to validate the data in f.* [4]

Then we can represent the next stage by the same function

$$issue: Forms \rightarrow Identifier.$$

These two stages are represented by composing the functions to make

$$generate: Forms \times Identifier_1 \times ... \times Identifier_k \rightarrow Identifier$$

where

$$generate(f, i_1, ... , i_k) = \textbf{if } check(f, i_1, ... , i_k)=1 \textbf{ then } issue (f) \textbf{ else } reject$$

## 9. Comparing identifiers

Consider the case where a set *Entity* of entities has two systems of identifiers:

$$IdSys_1 = (Entity, Identifier_1, id_1: Identifier_1 \rightarrow Entity),$$

$$IdSys_2 = (Entity, Identifier_2, id_2: Identifier_2 \rightarrow Entity).$$

How can we relate or compare these systems?

One simple case is when the identifiers in *Identifier_1* can be associated or matched with one or more identifiers in *Identifier_2*. This means that given an identifier $i \in Identifier_1$ of an entity

---

[4] It is easy to represent the case where the identifiers add information to that in the form: the function would have the form *issue (f,$i_1$, ... ,$i_k$)*.

*e*∈*Entity*, we can find *some* corresponding identifier in *Identifier$_2$* that is *also* an identifier for *e*. This is formalised as follows:

**Definition.** The system of identifiers *IdSys$_1$* is said to *reduce* to the system of identifiers *IdSys$_2$* if there is a single-valued mapping

$$red: Identifier_1 \rightarrow Identifier_2$$

that calculates for each identifier in *Identifier$_1$* a corresponding identifier in *Identifier$_2$* for entities in the following precise sense: for every entity *i*∈ *Identifier$_1$*,

$$id_1(i) = id_2(red(i))$$

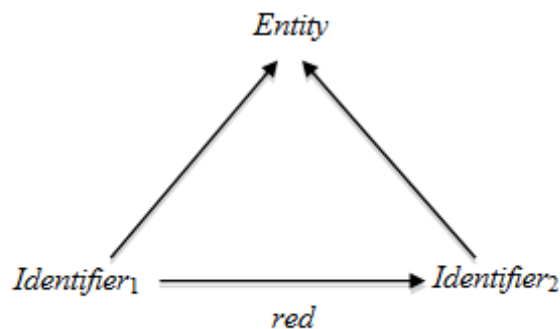

Figure 4: Reducing one system of identifiers to another

*Example.* Consider the set *Keep* of keepers of vehicles in the UK and two systems of identity for this set of entities. Suppose, for simplicity, *each keeper has one car*. Each car has a registration mark. Let the first system be

$$Reg = (Keep, Regmk \mid reg: Regmk \rightarrow Keep).$$

Every keeper has an address assigned by the postal service so let

$$Add = (Keep, Address \mid addr: Address \rightarrow Keep).$$

Then the Driver and Vehicle Licensing Agency (DVLA) is responsible for the determining the keeper's address from the registration mark, which is defined formally by the reduction map *red: Regmk→Address* such that for every registration mark r ∈*Regmk*,

$$reg(r) = addr(red(r)).$$

We say that the system of identities *Reg* is reducible to *Add*.

## 10. Conclusion

We have used formal methods to model precisely the main conceptual components involved in surveillance. To isolate and define ideas in great generality is the *raison d'être* of formal methods, though their mathematical nature presents obstacles to their reception and appreciation.

Our analysis of surveillance and identity has focused on technologies that collect and process data. At the heart of such technologies are software systems, which by their nature are examples of formalised systems. In surveillance, we study the representation of various forms of data (visual, audio and textual) and computation and communication with these data. Thus, the conceptual modelling of such systems is naturally and necessarily and ultimately a matter for mathematical formalisation. The formal framework we offer is both a rigorous analysis of the conceptual structure of surveillance and a starting point for technical questions about software.

Let us observe that increasingly our professional and social interaction is carried out – or controlled – by abstract technological systems rather than by direct face-to-face interactions. Whilst interacting with each of these systems, an individual needs to give over some of his/her identity to distinguish himself/herself from other users. Thus, rather than having a single and holistic identity, individuals now have many separated and overlapping identities. The multiplicity of identity, especially the extension of identity from the physical to the virtual world, requires the nature of identity needs to be problematised.

In the new era, the physical and the virtual are converging. In particular, the physical world is being sucked into the virtual, causing muddle and confusion. We have only recently begun to create a virtual world that is independent and shows signs of autonomy. Clearly, the physical and the virtual are fundamentally different and we must negotiate their co-existence. With this prospect, all the components and methods of surveillance – entity, observable behaviour, attribute and identity – will exist more definitively and naturally in their virtual forms, which are those of data and software. Thus, in practice, surveillance will increasingly become dominated by methods for the production, communication and publication of data and their scientific, social and political consequences.

Finally, let us observe that whenever a social science topic – in this case surveillance – is closely associated with technology – especially with technological tools that collect and process data effectively – then the specification of the software tools, i.e., what they do for users, can be formalised in much the same way as we have approached the problem here. Thus, the sociological notions that motivate, shape and are ultimately represented in the specification of software can be defined in a formal framework which can be mathematically analysed. In short,

**Principle.** *Sociological topics that are closely associated with software techniques can be expected to have formal theories.*

The use of formal methods to express and analyse general notions is a commonplace in areas of philosophy and linguistics but seems to be rare in social studies given the invasion of software into professional and social life, and the chain reaction on data, the role of formal methods to theorise about social concepts and problems is destined to grow.